\documentclass[twocolumn,showpacs,preprintnumbers,amsmath,aps]{revtex4}
\usepackage{graphicx}
\usepackage{dcolumn}
\usepackage{bm}
\begin{document}
\newcommand{\kvec}{\mbox{{\scriptsize {\bf k}}}}
\def\eq#1{(\ref{#1})}
\def\fig#1{\hspace{1mm}\ref{#1}}
\def\tab#1{table\hspace{1mm}\ref{#1}}
\title{Specific Heat and Thermodynamic Critical Field for Calcium under the Pressure at 120 GPa}
\author{R. Szcz{\c{e}}{\`s}niak, A.P. Durajski, M.W. Jarosik}
\affiliation{Institute of Physics, Cz{\c{e}}stochowa University of Technology, Al. Armii Krajowej 19, 42-200 
                Cz{\c{e}}stochowa, Poland}
\email{adurajski@wip.pcz.pl}
\date{\today} 
\begin{abstract}
The free energy difference between the superconducting and normal state for Calcium under the pressure at $120$ GPa has been determined. The numerical calculations have been made in the framework of the imaginary axis Eliashberg approach. On the basis of the obtained results the specific heat in the superconducting $C^{S}\left(T\right)$ and normal $C^{N}\left(T\right)$ state, as well as, the thermodynamic critical field $H_{C}\left(T\right)$ have been obtained. It has been shown that the characteristic values of the considered thermodynamic quantities do not obey the BCS universal laws. In particular, ${\Delta}C\left(T_{C}\right)/C^N\left(T_{C}\right)=2.48$ and $T_{C}C^{N}\left(T_{C}\right)/H_{C}^{2}\left(0\right)=0.154$.
\end{abstract}
\maketitle
%
\section{Introduction}

The superconducting properties of Calcium under the high pressure ($p$) are intensively studied since 1981 (Dunn and Bundy) \cite{Dunn}. The pressure dependence of the critical temperature ($T_{C}$) has been determined by Yabuuchi {\it et al.} in 2006 \cite{Yabuuchi}. The obtained results showed that $T_{C}$ increases significantly with the increasing pressure from $3$ K to $23$ K for $p\in\left(58,113\right)$ GPa. Above $113$ GPa the grow of $T_{C}$ is considerably slower and at $161$ GPa the critical temperature reaches the maximum value, which is equal to 25 K. We notice that in the considered pressure region, Calcium shows the complicated structural phase transitions \cite{Yabuuchi}-\cite{Nakamoto}. The proposed structural phase diagrams for Calcium the reader can find in \cite{Nakamoto1}.

In the presented paper we have calculated the free energy difference between the superconducting and normal state for Calcium under the pressure at $120$ GPa ($T_{C}=24$ K). Next, the specific heat and the thermodynamic critical field have been determined. The numerical analysis was based on the Eliashberg equations on the imaginary axis \cite{Eliashberg}. 

Let us pay attention that the Eliashberg approach extends the original idea of Bardeen, Cooper and Schrieffer \cite{BCS}, taking exactly into consideration the electron-phonon interaction. In the framework of the Eliashberg formalism, the strong coupling corrections to the BCS results are dependent on the value of the parameter $k_{B}T_{C}/\omega_{{\rm ln}}$. The symbol $\omega_{{\rm ln}}$ is called the logarithmic phonon frequency and $\omega_{{\rm ln}}\equiv \exp\left[\frac{2}{\lambda}\int^{\Omega_{\rm{max}}}_{0}d\Omega\frac{\alpha^{2}F\left(\Omega\right)}{\Omega}\ln\left(\Omega\right)\right]$. For Calcium, the Eliashberg function $(\alpha^2F(\Omega))$ has been calculated in the paper \cite{Yin}, the maximum phonon frequency ($\Omega_{{\rm max}}$) and the electron-phonon coupling constant ($\lambda$) are equal to $61.68$ meV and $1.3$ respectively. In the case of the BCS limit, the Eliashberg function is non-zero only for very high frequency, so that $k_{B}T_{C}/\omega_{{\rm ln}}\rightarrow 0$. 
In Calcium, value of the ratio $k_{B}T_{C}/\omega_{{\rm ln}}$ is equal to $0.082$. In this case the thermodynamic parameters can't be calculated exactly in the framework of the BCS model.

\section{THE ELIASHBERG EQUATIONS}

\begin{figure} [ht]
\includegraphics[scale=0.30]{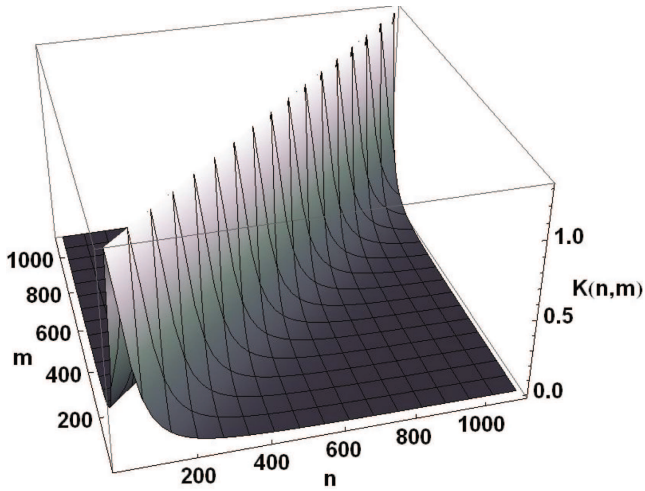}
\caption{\label{f1}
The pairing kernel $K\left(n,m\right)$ as a function of the numbers $n$ and $m$.}
\end{figure}

The Eliashberg equations on the imaginary axis can be written in the following form \cite{Eliashberg}: 
\begin{equation}
\label{r1}
\Delta_{n}Z_{n}=\frac{\pi}{\beta} \sum_{m=-M}^{M}
\frac{K\left(n,m\right)-\mu^{*}\theta\left(\omega_{c}-|\omega_{m}|\right)}{\sqrt{\omega_m^2+\Delta_m^2}}\Delta_{m} 
\end{equation}
and
\begin{equation}
\label{r2}
Z_n=1+\frac {\pi}{\beta\omega _n }\sum_{m=-M}^{M}\frac{K\left(n,m\right)}{\sqrt{\omega_m^2+\Delta_m^2}}\omega_{m},
\end{equation}
where the symbol $\Delta_{n}\equiv\Delta\left(i\omega_{n}\right)$ denotes the order parameter and $Z_{n}\equiv Z\left(i\omega_{n}\right)$ is the wave function renormalization factor; $n$-th Matsubara frequency is defined as: $\omega_{n}\equiv \frac{\pi}{\beta}\left(2n-1\right)$, where $\beta\equiv 1/k_{B}T$.
The electron-phonon pairing kernel $K\left(n,m\right)$ is given by:  
\begin{equation}
\label{r3}
K\left(n,m\right)\equiv 2\int_0^{\Omega_{\rm{max}}}d\Omega\frac{\Omega}
{\left(\omega_n-\omega_m\right)^2+\Omega ^2}\alpha^{2}F\left(\Omega\right).
\end{equation}
In Fig.\fig{f1} we have presented the form of $K\left(n,m\right)$ for the positive Matsubara frequencies and the temperature $1.16$ K. It is easy to notice that the pairing kernel is always positive and it achieves the strong maximum for $\omega_{n}=\omega_{m}$. The above result means, that Eqs. \eq{r1} and \eq{r2} can have the superconducting solution ($\Delta_{n}\neq 0)$.

%
%

In the framework of the Eliashberg formalism, the depairing electronic interaction is described by the Coulomb pseudopotential: 
${\mu^*}\equiv\mu\left[1+\mu\ln\left(\frac{\omega_{P}}{\omega_{D}}\right)\right]^{-1}$, where $\mu$ is defined by: $\mu\equiv\rho\left(0\right)V_{C}$. The symbol $\rho\left(0\right)$ denotes the value of the electronic density of states at the Fermi energy and $V_{C}$ is the Coulomb potential. The quantities $\omega_{P}$ and $\omega_{D}$ are the electronic plasma frequency and the Debye phonon frequency respectively \cite{Eliashberg}.We have calculated the value of the Coulomb pseudopotential in the paper \cite{Szczesniak1}. The following result has been obtained: ${\mu^*}=0.215$. Finally, $\Theta$ is the Heviside unit function and $\omega_{c}$ represents the cut-off frequency: $\omega_{c}=3\Omega_{\rm{max}}$. 

The Eliashberg equations have been solved for $2201$ Matsubara frequencies ($M=1100$) by using the method presented in \cite{Szczesniak2} and \cite{Szczesniak3}. In the considered case, the obtained Eliashberg solutions are stable for $T\geq 1.16$ K.

\section{The numerical results}

\begin{figure} [ht]
\includegraphics[scale=0.31]{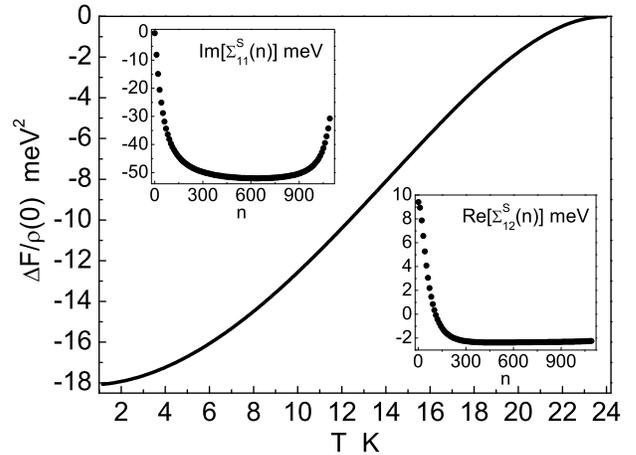}
\caption{\label{f2}
The free energy difference between the superconducting and normal state as a function of the temperature. The insets show the diagonal and non-diagonal elements of the matrix self energy on the imaginary axis.}
\end{figure}

The free energy difference between the superconducting and normal state ($\Delta F$) for an interacting electron-phonon systems should be determined by using the expression \cite{Bardeen}: 
\begin{eqnarray}
\label{r4}
\frac{\Delta F}{\rho\left(0\right)}&=&-\frac{2\pi}{\beta}\sum_{n=1}^{M}
\left(\sqrt{\omega^{2}_{n}+\Delta^{2}_{n}}- \left|\omega_{n}\right|\right)\\ \nonumber
&\times&(Z^{S}_{n}-Z^{N}_{n}\frac{\left|\omega_{n}\right|}
{\sqrt{\omega^{2}_{n}+\Delta^{2}_{n}}}),  
\end{eqnarray}  
where $Z^{S}_{n}$ and $Z^{N}_{n}$ denote the wave function renormalization factors for the superconducting ($S$) and normal ($N$) state respectively. 
In Fig.\fig{f2} we have plotted the dependence of ($\Delta F$) on the temperature. From the physical point of view, the negative values of $\Delta F$ prove that the superconducting state is stable to the critical temperature. 

%
%

We notice that the knowledge of the function $Z^{S}_{n}$ and $\Delta_{n}$ enables also the determination of the open form of the electronic self energy for the superconducting state ($\Sigma^{S}\left(n\right)$) \cite{Eliashberg}. In the framework of the Eliashberg formalism, $\Sigma^{S}\left(n\right)$ is represented by the $2\times 2$ matrix. For the half-filled energy band, the diagonal elements of $\Sigma^{S}\left(n\right)$ are imaginary and the non-diagonal elements are real. Additionally, it is occurring: $\Sigma^{S}_{11}\left(n\right)=\Sigma^{S}_{22}\left(n\right)$ and $\Sigma^{S}_{12}\left(n\right)=\Sigma^{S}_{21}\left(n\right)$.  
The dependence of Im$\left[\Sigma^{S}_{11}\left(n\right)\right]$ and Re$\left[\Sigma^{S}_{12}\left(n\right)\right]$ on the positive values of the number $n$ is presented in the insets in Fig.\fig{f2}. It is easy to see that the values of Im[$\Sigma^{S}_{11}\left(n\right)$] can be relatively high in comparison with Re[$\Sigma^{S}_{12}\left(n\right)$] and the considered function does not saturate for the large values of $n$. In the second case, we have found that Re[$\Sigma^{S}_{12}\left(n\right)$] has approximately a form of the Lorenz function.  

%
\begin{figure} [ht]
\includegraphics[scale=0.31]{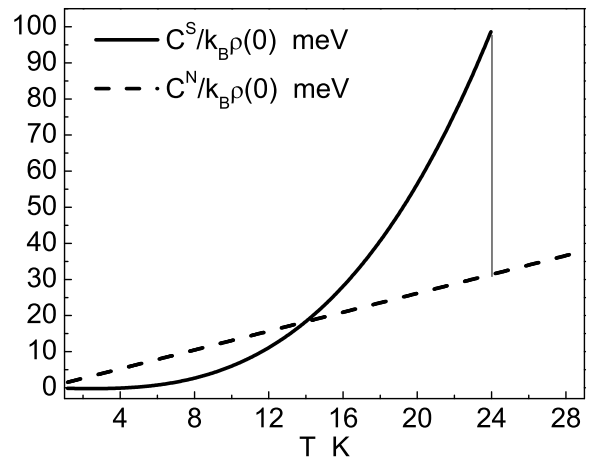}
\caption{\label{f3}
The specific heat for the superconducting and normal state as a function of the temperature.}
\end{figure}
%
\begin{figure} [ht]
\includegraphics[scale=0.31]{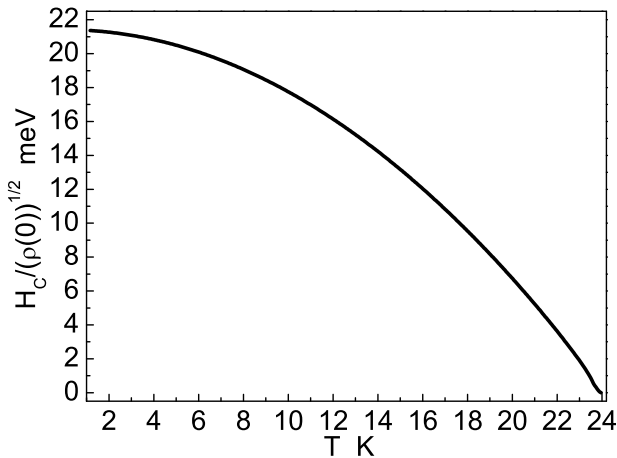}
\caption{\label{f4}
The thermodynamic critical field as a function of the temperature.}
\end{figure}
%

The difference in the specific heat between the superconducting and normal state $\left(\Delta C\equiv C^{\rm{S}}-C^{\rm{N}}\right)$ should be calculated by using the expression: $\frac{\Delta C\left(T\right)}{k_{B}\rho\left(0\right)}=-\frac{1}{\beta}\frac{d^{2}\left[\Delta F/\rho\left(0\right)\right]}{d\left(k_{B}T\right)^{2}}$. On the other hand, the specific heat in the normal state can be obtained with the help of the formula: $\frac{C^{N}\left(T\right)}{ k_{B}\rho\left(0\right)}=\frac{\gamma}{\beta}$, where the Sommerfeld constant is given by: $\gamma\equiv\frac{2}{3}\pi^{2}\left(1+\lambda\right)$. The dependencies of the specific heats on the temperature have been presented in Fig. \fig{f3}.  It is easy to see that at low temperatures, the specific heat in the superconducting state is exponentially suppressed. For the higher temperatures $C^ {S}$ rapidly increases and the values of the specific heat in the superconducting state much exceed the values of $C^{N}$. At the critical temperature the characteristic jump has been marked by the vertical line.

Below we have calculated the values of the thermodynamic critical field (cgs units):
$\frac{H_{C}}{\sqrt{\rho\left(0\right)}}=\sqrt{-8\pi\left[\Delta F/\rho\left(0\right)\right]}$.
The temperature dependence of $H_{C}/\sqrt{\rho\left(0\right)}$ has been shown in Fig.\fig{f4}.

The knowledge of the thermodynamic functions $C^{S}$, $C^{N}$ and $H_{C}$ enables the determination of the fundamental ratios:
\begin{equation}
\label{r5}
r_{1}\equiv\frac{\Delta C\left(T_{C}\right)}{C^{N}\left(T_{C}\right)} \qquad {\rm and} \qquad
r_{2}\equiv\frac{T_{C}C^{N}\left(T_{C}\right)}{H_{C}^{2}\left(0\right)},
\end{equation}
where $H_{C}\left(0\right)\simeq H_{C}\left(T=1.16 K\right)$. For Calcium under the pressure at $120$ GPa we have obtained: $r_{1}=2.48$ and $r_{2}=0.154$. Let us notice that the above results strongly differ from the canonical BCS predictions. In particular, $\left[r_{1}\right]_{{\rm BCS}}=1.43$ and $\left[r_{2}\right]_{{\rm BCS}}=0.168$ \cite{BCS}.
 
\section{Concluding Remarks}

The imaginary axis Eliashberg equations for Calcium under the pressure at $120$ GPa have been exactly solved in the paper. On the basis of the obtained results we have calculated the free energy difference between the superconducting and normal state by using the expression given by Bardeen and Stephen. Next, the specific heat for the superconducting and normal state, as well as, the critical field have been determined. It has been shown that the ratios between the characteristic values of the calculated thermodynamic functions strongly differ from the values predicted by the BCS model. In particular, 
${\Delta}C\left(T_{C}\right)/C^N\left(T_{C}\right)=2.48$ and $T_{C}C^{N}\left(T_{C}\right)/H_{C}^{2}\left(0\right)=0.154$.


\begin{acknowledgments}
The authors wish to thank K. Dzili{\'n}ski for the creation of the excellent working conditions. All numerical calculations were based on the Eliashberg function for Calcium sent to us by: Z.P. Yin and W.E. Pickett for whom we are also very thankful.
\end{acknowledgments}


%
\end{document}